\documentclass[sn-mathphys,iicol]{sn-jnl}
\usepackage{amsmath}
\usepackage{graphicx} 
\usepackage{mathrsfs}
\usepackage{amssymb}
\usepackage{lmodern}
\usepackage{mathtools}
\usepackage{hyperref}
\usepackage{subfig}
\usepackage{comment}
\usepackage{braket}
\providecommand{\abs}[1]{\lvert#1\rvert}  

\jyear{2022}%

\theoremstyle{thmstyleone}%
\theoremstyle{thmstyletwo}%

\theoremstyle{thmstylethree}%

\begin{document}

\title[WE of entangled SB oscillators]{Wehrl entropy of entangled Segal-Bargmann oscillators}

\author[1]{\fnm{David} \sur{Alonso López}}\email{dalons07@ucm.es}
\author[1,2]{\fnm{Jose A. R.} \sur{Cembranos}}\email{cembra@ucm.es}
\author*[1]{\fnm{David} \sur{Díaz-Guerra}}\email{ddiazgue@ucm.es}
\author[1]{\fnm{Andrés} \sur{Mínguez Sánchez}}\email{anming01@ucm.es}

\affil[1]{\orgdiv{Departamento de Física Teórica}, \orgname{Universidad Complutense de Madrid}, \orgaddress{\street{Plaza de Ciencias 1}, \city{Facultad de Ciencias Físicas}, \postcode{28040}, \state{Madrid},
\country{Spain}}}

\affil[2]{\orgdiv{Institute of Particle and Cosmos Physics (IPARCOS)}, \orgname{Universidad Complutense de Madrid}, \orgaddress{\street{Plaza de Ciencias 1}, \city{Facultad de Ciencias Físicas}, \postcode{28040}, \state{Madrid},
\country{Spain}}}

\abstract{In this manuscript we study the Wehrl entropy of entangled oscillators. This semiclassical entropy associated with the phase-space description of quantum mechanics
can be used for formulating uncertainty relations and for a quantification of entanglement. We focus on a system of two coupled oscillators described within its Segal-Bargmann space. This Hilbert space of holomorphic functions integrable with respect to a given Gaussian-like measure is particularly convenient to deal with harmonic oscillators. Indeed, the Stone-von Neumann theorem allows us to work in this space in a full correspondence with the ladder operators formalism. In addition, the Husimi pseudoprobability distribution is directly computed within the Segal-Bargmann formalism. Once we obtain the Husimi function, we analyze the Wehrl entropy and mutual information.}

\keywords{ Wehrl entropy, Segal-Bargmann, Entanglement, Husimi distribution, Mutual information}



\maketitle

\section{Introduction}
The harmonic oscillator is a system that characterizes the movement of a particle over a rest point. As simple as it is, small perturbations of a non-linear system can be described by simple oscillators.  Because of this, the composition of multiple oscillators is able to describe wave-like systems. This property allows harmonic oscillators to be one of the most used tools in fundamental physics, essential for theories like quantum field theory or condensed matter physics. When this system is studied through Quantum Mechanics, the quantization nature is shown explicitly \cite{diracPrinciples}. The quantum harmonic oscillator is characterized by certain levels or occupation numbers represented by positive integers.  The energy of the system is an archetypal example of a property quantized in discrete levels. In the most standardized Quantum Mechanics formalism, the harmonic oscillator is described by creation and annihilation operators. These operators determine a Hilbert space, Fock space, characterized by the oscillator occupation number. Creation and annihilation operators work by raising or lowering the occupation level of the harmonic oscillator.

By using non-commutative operators for harmonic oscillators, standard quantum mechanics imposes the well-known commutative relations for the
creation and annihilation operators. Moreover, quantum mechanics uses dynamical functions living typically in the configuration space to describe a system. On the contrary, the most fundamental space in classical physics is the phase space, where position and momentum live in equal footing. The incompatibility between quantum operators and classical functions can be solved with the development of a phase space quantum mechanics formalism \cite{zachos}. The most popular approach to this task is using pseudo-probability distributions like Wigner function or Husimi function \cite{wigner, weyl, husimi, lee95}, both defined in phase space.  This quantum mechanics formalism is interesting in relation to the symplectic formalism associated with classical mechanics.

Describing harmonic oscillators with these distributions is not the only way of representing quantum harmonic oscillator with phase-space functions and conserving non-commutativity at the same time. Segal-Bargmann Space (SBS) is a Hilbert space span by holomorphic functions, that preserves non-commutativity by including a Lebesgue measure \cite{segal,bargmann61, hall00}. This space allows a description of quantum systems in terms of holomorphic functions. In addition, the Husimi function can be calculated directly from SBS functions, acting as a bridge between both descriptions.

One of the quantum phenomena that mostly challenged the preconception of physics when quantum mechanics was discovered is quantum entanglement. Quantum entanglement \cite{horodecki} is a consequence of two fundamental properties of quantum mechanics: the superposition principle and the composition of multiple systems, so it is an inevitable quantum feature. In this phenomenon, two (or multiple) systems that once interacted start sharing certain correlations in quantum measurements. This means that measurements in one system are correlated with the other one with no need of an interaction between them when the measurement takes place. This caused some problems in how reality was conceived in quantum mechanics \cite{epr}, such as the causality paradox and its relation to the locality principle \cite{bell}. 

SBS allows to naturally study the behavior of harmonic oscillators using holomorphic functions. The superposition of multiple Hilbert spaces for multiple harmonic oscillators give rise to the possibility of quantum entanglement. The exploration of this entanglement has been carried out before from a theoretical and an experimental approach \cite{horodecki, makarov18, vedral03, NozCoupled}. However, the tools from SBS shows the consequences of quantum entanglement in phase space as well as different quantum properties like entropy or mutual information in a collection of quantum harmonic oscillators.

In this work, we study quantum harmonic oscillators by the composition of independent SBSs. This formalism allows us to approach the Wehrl entropy of entangled oscillators by using holomorphic functions. 
Through a convenient basis change, we describe a system of independent harmonic oscillators with a coupling term that generates entanglement. From this state, we can define excited states by applying creator operators. From these states we can calculate different quantum properties like uncertainty relations, the purity of the system or the associated entropy. From our approach, it is straightforward to check that the Heisenberg uncertainty principle is fulfilled, as it must be  independently of the quantum mechanics description used. We show how a larger coupling implies a larger uncertainty threshold. It means that information from the individual subsystems is lost due to this coupling. We compute the Husimi quasiprobability distribution of the system directly from the Segal-Bargmann functions. This allows us to represent the probability density in the composed phase space of different subsystems. We also obtain the Wehrl entropy, a semiclassical entropy associated with Husimi function. We show that such entropy increases with the amount of entanglement of the system. We study all of these quantum properties for two coupled harmonic oscillators.

This manuscript is divided as follows. In Section \ref{sec:teo} we survey the theoretical fundamental properties of the SBS, which includes the orthonormal function basis and how it is related to a representation of creation and annihilation operators within the space, allowing an analysis of harmonic oscillators.  In Section \ref{sec:OA}, we describe a system formed by two coupled oscillators in the Segal-Bargmann formalism and its associated Husimi function. In Section \ref{sec:entropy}, we carry out the computation of the Wehrl entropy of the system for different levels of the complete system. Throughout the work, we use natural units with $k_B=\hbar=1$ unless explicitly stated.

\section{Segal-Bargmann formalism for harmonic oscillators}\label{sec:teo}

\subsection{The Segal-Bargmann space}

The SBS is defined as the space of square-integrable holomorphic functions on $U\subset\mathbb{C}^d$ with respect to the weight $\mu_\hbar$ that satisfies
\begin{align}
    \mathcal{H}L^2 &(U,\mu_\hbar) \equiv \biggl\{ F\in \mathcal{H}(U) \, \bigg \vert  \nonumber \\
    &   \bigg \vert  \, \| F \|^2 = \int_U \lvert F(z)\rvert^2 \mu_\hbar(z) \, \mathrm{d}z < \infty  \biggr\}.
\end{align}
Here $\mathcal{H}(U)$ denotes the space of holomorphic functions on $U$, $\mathrm{d}z$ is the $2d$-dimensional Lebesgue measure on $\mathbb{C}^d$ and $\mu_\hbar(z) = (\pi\hbar)^{-d}\,\text{exp}\{-\lvert z \rvert^2/\hbar\}$ is a continuous, strictly positive function on $U$ where $\lvert z \rvert^2 = \lvert z_1 \rvert^2 + \cdots + \lvert z_d \rvert^2$ and $\hbar$ is a positive number that for physical cases coincides with Planck constant \cite{segal, bargmann61, hall00}. Taking into account that the SBS has its own inner product, an orthonormal basis for this space acquires the expression
\begin{equation}
    \left\{\prod_{k=1}^d \frac{z_k^{m_k}}{\sqrt{\hbar^{m_k}\, m_k!}}\, : \, m_1,\cdots,m_d \in \mathbb{N}\right\}. 
    \label{Base_SBS}
\end{equation}

As a result of being a space of square-integrable holomorphic functions, the SBS has associated the basic properties: 
\begin{enumerate}
    \item Continuous pointwise evaluation: 
    \begin{align}
        &\exists c_z : \lvert F(z)\rvert^2 \leq c_z \, \lvert\lvert F\rvert\rvert_{L^2(U,\mu_\hbar)}^2 , \nonumber \\
        &\text{for all} \; F \in \mathcal{H}L^2(U,\mu_\hbar), \, \forall z\in U .
    \end{align}
    \item $\mathcal{H}L^2(U,\mu_\hbar)$ is a closed subspace of $L^2(U,\mu_\hbar)$, and hence a Hilbert space.
\end{enumerate}
Furthermore, in the SBS, there is a special function known as the reproducing kernel $K(z,w)$, whose expression can be obtained through an orthonormal basis, that has the following properties:
\renewcommand{\theenumi}{\roman{enumi}}
\begin{enumerate}
    \item It is holomorphic in $z$ and anti-holomorphic in $w$.
    \item It is a square-integrable function such that $\forall F \in \mathcal{H}L^2(U,\mu_\hbar)$ satisfies
    \begin{equation}
        F(z) = \int_U K(z,w) F(w) \, \mu_\hbar(w) \, \mathrm{d}w\,.
        \label{eq: s2e3}
    \end{equation}
    \item For all $z\in U$ it holds that $\lvert F(z)\rvert^2 \leq K(z,z)\, \lvert\lvert F(z)\rvert\rvert^2$.
\end{enumerate}
The mathematical expression for the reproducing kernel is
\begin{equation}
    K(z,w)=e^{z\cdot w^*/\hbar}\,,
    \label{Kernel_SBS}
\end{equation}
where $z\cdot w^* = z_1 w_1^* + \cdots + z_d w_d^*$ and the symbol $*$ above $w$ indicates complex conjugate. To obtain this expression, basis (\ref{Base_SBS}) was needed.\\ 

For simplicity, in the remainder of this manuscript we will consider that the subset $U$ is the whole $\mathbb{C}^d$ \cite{hall00} and that all inner products are assumed to be with respect to the $\mathcal{H}L^2(\mathbb{C}^d,\mu_\hbar)$ space. Also, when we do not specify the domain of integration, it is assumed that the integration is over the entire complex plane. Since the measure $\mathrm{d}z$ covers all $\mathbb{C}^d$ space, it then can be separated into its real and imaginary parts.

\subsubsection{Correspondence between $L^2(\mathbb{R}^{d},\mathrm{d}x)$ and $\mathcal{H}L^{2}(\mathbb{C}^{d},\mu_\hbar)$}

Let $z_k$ and $\hbar \, \partial/\partial z_k \equiv \hbar \partial_{z_k}$ be a pair of operators. Using an arbitrary holomorphic function $f \in \mathcal{H}(\mathbb{C}^d)$ it can be shown easily that
\begin{equation}
    \left[ \hbar \partial_{z_k}, z_l \right]f(z) = \hbar \, \delta_{kl} \, f(z)\,.
    \label{CCR}
\end{equation}
This expression may result familiar, since it looks a lot like the canonical commutation relationships (CCR) of the usual creation and annihilation operators, except for a factor $\hbar$. However, even though $z_k$ and $\hbar\partial_{z_k}$ have the same CCR as $a_k^{\dagger}$ and $a_k$ operators, they are not a representation of these. To establish a correct relationship between them, the SBS will be needed since it is a Hilbert space on which $z_k$ and $\hbar\partial_k$ act continuously, irreducibly and are adjoint to each other with respect to the inner product.\\

In addition, since $z_k$ and $\hbar\partial_{z_k}$ satisfy the Weyl algebra, the existence of a map between $L^2(\mathbb{R}^d,\mathrm{d}x)$ and $\mathcal{H}L^2(\mathbb{C}^d,\mu_\hbar)$ is guaranteed thanks to the Stone-von Neumann theorem. This map is known as the Segal-Bargmann transform \cite{segal, bargmann61, hall00} and it is defined as
\begin{equation}
    \begin{split}
        A_\hbar: L^2(\mathbb{R}^d,\mathrm{d}x) & \longrightarrow \mathcal{H}L^2(\mathbb{C}^d,\mu_\hbar) \\ 
        f(x) & \longmapsto A_\hbar f(z),
    \end{split}
    \label{TransformadaSB}
\end{equation}
with,
\begin{align}
    A_\hbar & f(z) =  (\pi\hbar)^{-d/4} \nonumber \\
    & \cdot \int_{\mathbb{R}^d} e^{(-z^2+2\sqrt{2}x\cdot z-x^2)/2\hbar}f(x)\mathrm{d}^dx.
\end{align}
Here $A_\hbar f$ is the equivalent to the wave function in the Schrödinger picture. The Segal-Bargmann transform satisfies the following properties: (i) The integral is convergent and it is a holomorphic function for all $f\in L^{2}(\mathbb{R}^d, \mathrm{d}x)$. (ii) The map $A_{\hbar}$ is a unitary map. (iii) For $k=1, \cdots, d$
\begin{equation}
    \begin{split}
        A_\hbar a_k A_\hbar^{-1} &= \hbar\frac{\partial}{\partial z_k}, \\
        \qquad A_\hbar a_k^{\dagger} A_\hbar^{-1} &= z_k\,.
    \end{split}
    \label{correspondenciaSB}
\end{equation}
Due to this last equivalence, in the SBS, it will be possible to rewrite the hermitian conjugate of $z_k$ operator in the following form
\begin{equation}
    \overline{z}_k=\hbar\partial_{z_k}\,.
\end{equation}
As a consequence, when we deal with the hermitian conjugate of $z_k$ we will use $\overline{z}_k$ and $\hbar\partial_{z_k}$ interchangeably.\\

An important detail to keep in mind are the physical units of the presented equations. Assuming that the operators in (\ref{CCR}) have the same units, then it is easy to see that $z_k$ and $\hbar\partial_{z_k}$ have units of $[ \text{Action} ]^{1/2}$. Similarly, using the expression of the Segal-Bargmann transform, the configuration variables $x_k$ from the Hilbert space $L^2(\mathbb{R}^d,\mathrm{d}x)$ must have the same units as $z_k$. This means that when we take natural units, all these variables are going to become dimensionless.

\subsubsection{The Husimi function from the Segal-Bargmann formalism}
The SBS represents a valuable tool for quantum mechanics in phase space since the normal and anti-normal ordinations of creation and annihilation operators are expressed naturally using the variables: $z$ and $z^*$. Another advantage of the SBS is that the Husimi function arises naturally in this space. Given a normalized function $f\in\mathcal{H}L^2(\mathbb{C}^d,\mu_\hbar)$ such that $\| f \|^2 = 1$, the Husimi function can be written as 
\begin{equation}
	F^H (z,z^*) = \frac{1}{(\pi\hbar)^d}\lvert f(z,z^*)\rvert^2 e^{-\lvert z\rvert^2}.
	\label{fun_Husimi}
\end{equation}
From this definition is easy to verify that when the Husimi function is integrated to the whole complex plane $\mathbb{C}^d$, its result is normalized to unity
\begin{equation}
	\int_{\mathbb{C}^d} F^H (z,z^*)\, \mathrm{d}z = 1.
\end{equation}

\subsection{The harmonic oscillator}
In this paper, we are interested in applying the Segal-Bargmann formalism to a system of harmonic oscillators. The resolution of the harmonic oscillator within the SBS is simple and their results are very useful as we will see in following sections.\\

Let $H$ be the Hamiltonian of a $d$-dimensional harmonic oscillator in the $L^2(\mathbb{R}^d,\mathrm{d}x)$ space. Its expression in the SBS is obtained through the Segal-Bargmann transforms at it follows
\begin{align}
    H_{SB}&=A_\hbar H A_\hbar^{-1} = \sum_{k=1}^d H_{SB}^{(k)} \nonumber \\
    &= \sum_{k=1}^d \omega_k \left( z_k\frac{\partial}{\partial z_k} + \frac{1}{2} \right)\,.
    \label{Ham_Oscilador}
\end{align}
Here $\omega_k$ represents the oscillation frequency of the k-th oscillator. Since the problem is separable, we only have to solve one oscillator to solve the whole problem. Solving the eigenvalue problem for a hamiltonian of the type $H_{SB}^{(k)}$ is pretty straightforward since the solutions must be proportional to $z_k^{n_k}$, where $n_k \in \mathbb{N}$ represents the level of the k-th oscillator. To find the normalization constant, we will use the inner product of the SBS, so
\begin{align}
        \lvert\lvert z_k^{n_k} \rvert\rvert^2 &= \int_{\mathbb{C}^d} (z_k^{n_k})^* z_k^{n_k} \, \mu_\hbar \, \mathrm{d}z \nonumber \\
        &= \frac{1}{\pi} \int_0^{2\pi} \mathrm{d}\theta_k \int_0^\infty r_k^{2n_k + 1} e^{-r_k^2}\, \mathrm{d}r_k \nonumber\\
        &= n_k!\,. 
\end{align}
Therefore, the eigenstates and the eigenvalues of the Hamiltonian $H_{SB}^{(k)}$ can be written as
\begin{equation}
	f_{n_k}(z_k) = \frac{z_k^{n_k}}{\sqrt{n_k!}}\,,
	\label{solSBOA}
\end{equation}
and $\epsilon_{n_k}=\omega_k(n_k + 1/2)$.
From these results, the eigenstates and the eigenvalues of the original problem can be written as
\begin{equation}
	\prod_{k=1}^d f_{n_k}(z_k)\,, \qquad E=\sum_{k=1}^d\epsilon_{n_k}\,.
    \label{solSBOA2}
\end{equation}
Notice that the basis (\ref{Base_SBS}) for the SBS, coincides with the eigenstates values of the problem. This fact establishes a direct relationship between $\mathcal{H}L^2(\mathbb{C}^d,\mu_\hbar)$ and the harmonic oscillator.

\subsubsection{Occupation number}
Through the use of the creation and annihilation operators, the occupation number operator is defined as $N_k = a_k^{\dagger} a_k \mapsto N_{SB}^{(k)} = z_k \partial_{z_k}$. The action of this operator on eigenstates (\ref{solSBOA}) is straightforward $N_{SB}^{(k)}\,f_{n_k}(z_k) = n_k\,f_{n_k}(z_k)$. So the expected value of the occupation number for a general eigenstate (\ref{solSBOA2}) is directly
\begin{align}
    \langle N_{SB}^{(k)} \rangle &=  \frac{1}{\pi}\int_{\mathbb{C}} f_{n_k}^* n_k f_{n_k} e^{-\lvert z_k \rvert^2} \mathrm{d}z_k = n_k\,.
\end{align}

\subsubsection{Husimi function for the harmonic oscillator}
Once we have calculated the solutions of the harmonic oscillator, we can compute its pseoudprobability density. Only a single harmonic oscillator will be used to keep the procedure simple, extending this expression to $N$ oscillators is done by composition of individual Husimi functions. Using Equation \eqref{fun_Husimi} for the one-dimensional case, the probability density reads
\begin{equation}
	F^H_n (x,p) = \frac{1}{\pi n!} \left( \frac{x^2 +p^2}{2} \right) ^n \exp \left\lbrace -\frac{x^2 + p^2}{2} \right\rbrace\,.
\end{equation}
In this expression we have switched to phase space coordinates using the fact that the configuration variable can be rewritten as $z = (x -i p)/\sqrt{2}$.\\

In the ground state $(n=0)$, the Husimi function takes the form of a Gaussian distribution located at the phase space origin.
\begin{equation}
    F^H_0 (x,p) = \frac{1}{\pi}  \exp \left\lbrace -\frac{x^2 + p^2}{2} \right\rbrace\,. \label{husimiGS}
\end{equation}
As we stated before, Husimi functions represent a pseudo-probability distribution on phase space, so for the ground state we see how the higher probability is in the origin. If we keep going for excited states $(n>0)$, the Husimi function starts forming rings damped by a Gaussian distribution around the origin. Recall that the solutions for the classical harmonic oscillator describe ellipses, that with a proper normalization become a circunference, in phase space. So, we see that the quantum behavior fits with the classical description of the problem.

\subsubsection{Infinite oscillators formalism}\label{sec_infty}
Expanding the Segal-Bargmann formalism to infinite dimension is straightforward. For this formalism we will need an infinite set of variables $\lbrace z_k\rbrace$ that are square-summable such that $\sum_{k=1}^{\infty} \abs{z_k}^2 < \infty$. With this set we can define a SBS of infinite dimension \cite{klauder}, where the closed measure is defined by
\begin{equation}
    \mathrm{d}\mu=\prod_{k=1}^{\infty}\frac{e^{-\abs{z_k}^2}}{\pi} \, \mathrm{d}z_k\,.
\end{equation}
Once we set this space, it is possible to generalize an arbitrary scalar field using the creation and annihilation operators from the SBS as
\begin{equation}
    \psi (x) = \sum_{k=1}^{\infty} f_k(x) z_k + f^*_k(x) \bar{z}_k \,.
    \label{psi}
\end{equation}
Here $x$ is a variable in a $s$-dimensional configuration space, $\lbrace f_k(x) \rbrace$ is a set of orthonormal functions called modes that satisfy $\int f_k(x) f_l^*(x)\,\mathrm{d}x = \delta_{kl}$. Since $[\bar z_k , z_l] = \delta_{kl}$ is fulfilled then $\psi(x)$ and its canonical conjugate momentum fulfill their CCR. In addition, there must exist a vacuum state, which we will call $\Omega_z$, which satisfies that for all values of $k$ that it is annihilated by the annihilation operators. These definitions are enough for working with a field theory with operators, as Segal-Bargmann variables act as a representation of creation and annihilation operators \cite{hall00}.\\

We can expand variables from one Segal-Bargmann space into functions of another Segal-Bargmann space $\{w_k\}$ using Bogoliubov transformations
\begin{equation}
	w_j = \sum_{k=1}^{\infty} a_{kj}  \bar z_k + b_{kj}  z_k\,.
\end{equation}
Notice that all these operators must obey the commutation relations $[\bar w_i, w_j] = \delta_{ij}$ and $[w_i, w_k] =[\bar w_i, \bar w_k]=0$. So the coefficients are constrained by
\begin{equation}
	\sum_{k=1}^{\infty} (a_{ij} a_{jk}^* - b_{ij} b_{jk}^*) = \delta_{ij}, \quad \sum_{k=1}^{\infty} ( a_{ik} b_{jk} - b_{ik} a_{jk})= 0 \,.
\end{equation}
Notice also that this new set of operators $\lbrace w_k \rbrace$ has a different vacuum state than the one from the set $\lbrace z_k \rbrace$. In other words, the new vacuum state has to satisfy $\bar w_k \Omega_w = 0$ for all values of $k$. For set completeness, the scalar field $\psi (x)$ from equation (\ref{psi}) can be rewritten using the new set of operator $\lbrace w_k \rbrace$ and its modes $\lbrace g_k(x) \rbrace$ as  
\begin{equation}
	\psi (x) = \sum_{k=1}^{\infty} g_k(x) w_k + g^*_k(x) \bar{w}_k \, .
\end{equation}
Thus, there is arbitrariness when choosing a set of modes $\{f_k(x)\}$ or $\{g_k(x)\}$ that characterizes the field. This leads to the conclusion that different observers may define the same field with different operators, that is, using a different SBS.

\section{Quantum entanglement in phase space}\label{sec:OA}

\subsection{Entanglement in Hilbert spaces $\mathcal{H}$}
Quantum entanglement is a consequence of two fundamental principles of quantum mechanics: the superposition principle and the composition of Hilbert spaces. A composed system that is in a Hilbert space $\mathcal{H}$, can be expressed as a direct product between $m$ subsystems in their Hilbert subspaces $\mathcal{H}_i$ \cite{horodecki},
\begin{equation}
	\mathcal{H} = \otimes_{i=1}^{m} \mathcal{H}_i.
\end{equation}

The superposition principle ensures us that any state $\lvert \psi \rangle \in \mathcal{H}$ can be written as 
\begin{equation}
	\lvert \psi \rangle= \sum_{[j_m]} c_{[j_m]} \lvert [e_{j_m}]\rangle,
\end{equation}
$[j_m]= j_1,\cdots, j_m$ being a set of indices, where each $j_i$ is associated with its corresponding subsystem. Keeping in mind that each subsystem $\mathcal{H}_i$ has a basis $\lbrace \lvert e_{j_i} \rangle \rbrace_{j_i=1}^{d_i}$ where $d_i = \text{dim}\mathcal{H}_i$, then $\lvert [e_{j_m}]\rangle = \lvert e_{j_1}\rangle \otimes \cdots \otimes \lvert e_{j_m} \rangle$ represents an element of the basis that describes the whole system $\mathcal{H}$.\\
Depending on the values that the coefficients $\lbrace c_{[j_m]} \rbrace$ take, we can classify the states of $\mathcal{H}$ into two types. We will call separable states, to those states that can be written as the direct product of states of the $m$ subsystems, $\lvert \psi\rangle = \lvert \psi_1\rangle \otimes \cdots \otimes \lvert \psi_m\rangle$. If this is not the case, the state is said to be entangled.\\
For the phase space analysis, it is important to know how entanglement is identified in terms of functions. Given the representation of these states in the space of a set of variables $\{\eta\}$, 
\begin{equation}
	\psi ( \eta )  =  \sum_{[i_m]}c_{[i_m]} [\psi_{[i_m]}( \eta_i ) ],
\end{equation}
with $[\psi_{[i_m]}(\eta_i)] \equiv \psi_A (\eta_i) \otimes \cdots$. If $\psi(\eta)$ can be separated in functions for each variable $\eta_i$, then the system is separable, otherwise the system is entangled. This property can be applied to our functions in the SBS.\\

The definition of quantum entanglement can be extended to mixed states. Now, for a given composed system, each subsystem $\mathcal{H}_A, \mathcal{H}_B,\dots$ has a density operator $\rho_A, \rho_B,\dots$ associated. Then, we will say that a mixed state $\rho \in \mathcal{H}$ is separable if its expression can be written as a weighted sum of the tensor product of the densisty operators for each subsystem,
\begin{equation}
	\rho = \sum_{i} \omega_i \rho_i^{A} \otimes \rho_i ^{B} \otimes \cdots \, .
\end{equation}
The weights $\omega_i$ must satisfy the condition of being normalized $\sum_{i} \omega_i = 1$, and being positive $\omega_i \geq 0$ for all values of $i$. Equivalently to a pure state, a mixed state is entangled if it cannot be expressed in this way.

\subsection{Coupled harmonic oscillators}

For this section, we study the case of two coupled harmonic oscillators, which are well known to be entangled, in Segal-Bargmann formalism. Within this context, we use the tools provided by this space and the straightforward pseudoprobability density in phase space: the Husimi function. Some studies of coupled harmonic oscillators have been done with the second quantization formalism \cite{vedral03} and with the Wigner function \cite{NozCoupled}.

Let the Hamiltonian for two coupled one-dimensional harmonic systems expressed in their respective creation and annihilation operators in Segal-Bargmann space be,
\begin{equation}
    H_{SB} (z_1,z_2) = \omega \left ( z_1 \bar z_1  +z_2 \bar z_2 \right) + \lambda ( z_1 z_2 + \bar z_1 \bar z_2) + H_0.
\end{equation}
with $H_0$ a constant associated with the energy minimum. We call this structure conformed by two coupled oscillators: $\mathcal{Z}$.

Analytically solving the eigenvalue equation $H_{SB} f(z) = E f(z)$ for the time-independent energy is complicated as it has a non-linear term. For solving the equation is useful to perform Bogoliubov transformations \cite{bogo} that diagonalize the Hamiltonian to another set of creation and annihilation operators. Thus, a new set of ladder operators are defined in a new Segal-Bargmann space that characterize a new system, named $\mathcal{W}$.  We will work with two sets of variables: $w_1$, $\bar w_1$, $w_2$, and $\bar w_2$; that are related with the variables of $\mathcal{Z}$ via the particular Bogoliubov transformation given by,
\begin{align}
	z_1& = a_1 w_1 + b_2 \bar w_2, \\
	z_2 &= a_2 w_2 + b_1 \bar w_1 .
\end{align}

This new space is described by the same properties discussed in the last section and it is, at first sight, different than the coupled system $\mathcal{Z}$. Thus, they must satisfy the canonical commutation relations for ladder operators in Segal-Bargmann: $[\bar w_i, w_i] = 1$. Explicitly,
\begin{align}
	[\bar z_1 , z_1 ] &= \lvert a_1 \rvert ^2 - \lvert b_2 \rvert^2 = 1,\label{cond_bogo1} \\
	[\bar z_2, z_2 ] &= \lvert a_2 \rvert^2 - \lvert b_1 \rvert^2 = 1. \label{cond_bogo2}
\end{align}
A solution for these conditions is to set $a_1 = a_2 = \cosh \eta$ and $b_1 =b_2= \sinh \eta$, with $\eta$ a parameter. The last conditions \eqref{cond_bogo1} and \eqref{cond_bogo2} allow us to introduce a phase factor $e^{i \alpha_i}$ as the coefficient conditions are phase invariant. For simplicity, we set this phase to $\alpha=0$, but it can influence the transformations as seen by Noz et al. \cite{NozCoupled}.

Then, the full transformations that relate both systems depending on the $\eta$ parameter read,
\begin{align}
	w_1& = \cosh\eta\; z_1 - \sinh \eta\; \bar z_2, \notag \\
	w_2 &= \cosh \eta z_2\; - \sinh\eta\; \bar z_1 . \label{op_w}
\end{align} 

Then, the diagonalized Hamiltonian in the $\mathcal{W}$ system is
\begin{equation}
    \mathcal{H}(w_1,w_2) = \omega' \left(w_1 \bar w_1 + w_2 \bar w_2 \right) + H'_0,
\end{equation}
where the diagonalization imposes that the non-diagonal terms correspondent to $w_1$, $w_2$, and $\bar w_1$, and $\bar w_2$ have to be zero. This condition establishes $\lambda=-\omega\, \tanh(2\eta)$, which sets $\eta$ as the parameter that describes the coupling between oscillators. When $\eta = 0$, the coupled system $\mathcal{Z}$, it is already diagonal so in this case $\mathcal{Z}$ and $\mathcal{W}$ are equivalent. In this new Hamiltonian,  $\omega' \equiv \omega\, \text{sech}(2\eta)$ and $H'_0\equiv \omega' - \omega + H_0$, are defined as the uncoupled energy and energy minimum, respectively.

The coupled harmonic oscillator functions' space corresponding to the $\mathcal{Z}$ system is obtained by the successive application of creation operators on the ground or vacuum state of such a system $\Omega_z$. Thus, the vacuum states satisfies,
\begin{equation}
	\bar z_i \Omega_z = \partial_{z_i} 1 =  0, \quad \forall i.
\end{equation}
For the SBS corresponding to $\mathcal{Z}$ the measure is
\begin{equation}
	\mu_{z} = \frac{1}{\pi^2}\exp \left( -\lvert z_1 \rvert^2 - \lvert z_2\rvert^2\right) \label{espacio_z}
\end{equation}

The SBS for the $w_i$ variables can be characterized by defining the ground state $\Omega_w$, so that it is annhilated by $\bar{w}_i$ $\forall\,  i$:
\begin{equation}
    \bar w_i \Omega_w = \partial_{w_i} \Omega_w = 0.
\end{equation}

The excited states in the $\mathcal{W}$ decoupled system are generated by succesive application of the \eqref{op_w} operators to the ground state $\Omega_w$. 
The explicit form of the ground state of $\mathcal{W}$ in terms of the $z_i$ variables can be calculated by setting:
\begin{align}
	\bar w_1 \Omega_w&= (  \cosh\eta\;  \bar z_1 - \sinh \eta\; z_2)\, \Omega_w = 0, \\
	\bar w_2 \Omega_w & = ( \cosh \eta\; \bar z_2 - \sinh \eta\;  z_1 )\, \Omega_w = 0.
\end{align}
By using the method from \cite{shanta}, one can find the ground state to be
\begin{equation}
	\Omega_w (z) \propto \exp \left( \tanh \eta\; z_1 z_2 \right)\, \Omega_z,
\end{equation}
for the space defined by \eqref{espacio_z}, which is analogous to that of a harmonic oscillator in thermal equilibrium \cite{NozCoupled}. To obtain the normalization constant, we impose the following relation:
\begin{equation}
    \langle 1 \rangle_{\Omega_w} = \abs{\mathcal{N}}^2 \int \mu_z\,  e^{ \tanh \eta\; (z_1 z_2 + z_1^* z_2^*)}\, \mathrm{d} z_1  \mathrm{d}z_2 = 1. 
\end{equation}
Setting the real and imaginary parts $u_i$ and $v_i$ respectively so that $z_i = u_i + i v_i$,
\begin{align}
	\langle 1 \rangle_{ \Omega_w}  &= \frac{\abs{\mathcal{N}}^2}{\pi^2}\int \mathrm{d}u_1 \mathrm{d}u_2 \mathrm{d}v_1 \mathrm{d}v_2 \nonumber \\
	& \qquad \qquad \cdot  e^{ -u_1^2 - v_1^2 -u_2^2 - v_2^2} e^{2 \tanh \eta\; ( u_1 u_2 - v_1 v_2)} \notag \\
	&= \int \frac{\abs{\mathcal{N}}^2}{\pi} e^{-\text{sech}^2\eta\; \left(u_1^2+v_1^2\right)}\, \mathrm{d}u_1  \mathrm{d}v_1 \nonumber \\
	& =\abs{\mathcal{N}}^2 \cosh^2 \eta = 1,
	\label{normalización}
\end{align}

implies that $\mathcal{N}^{-1}= \cosh \eta$. Now we have $\mathcal{W}$ fully characterized by the ground state,
\begin{equation}
    \Omega_w (z) = \frac{1}{\cosh \eta}\, e^{  \tanh \eta\; z_1 z_2}, \label{vacio_w}
\end{equation}
from which we can compute the form of any desired excited state. The most general form of these states is:

\begin{equation}
	f(w)=\frac{1}{\sqrt{n_1!n_2!}}\,w_1^{n_1}w_2^{n_2}\,\Omega_w.
	\label{excited}
\end{equation}

Note that even the ground state \eqref{vacio_w} is clearly entangled since it is not possible to factorize the $z_1$ dependence from the $z_2$ dependence. The states given in \eqref{excited} are also entangled as they are obtained from the entangled ground state.  For this reason, any measurement performed upon this system will exhibit an entangled behaviour.

\subsubsection{Entanglement of the ground state}

As stated before, the ground state of the $\mathcal{W}$ system is entangled when expressed in terms of the variables $z_i$ of the $\mathcal{Z}$ system. It is convenient to analyze its entanglement from its associated Husimi function:

\begin{equation}
	F^H(z_1,z_2) = \frac{1}{\pi^2} e^{-\abs{z_1}^2 - \abs{z_2}^2 } \abs{f(z)}^2.
\end{equation}
In particular, for the ground state $\Omega_w$, we get:
\begin{align}
	F^H_{\Omega_w} &=  \frac{\text{sech}^2 \eta}{\pi^2} e^{-\abs{z_1}^2 - \abs{z_2}^2 } e^{ \tanh \eta\, ( z_1 z_2 + z^*_1 z^*_2)} \nonumber\\ = &\frac{\text{sech}^2 \eta}{\pi^2} e^{ - u_1^2 - v_1^2 } e^{-u_2^2 - v_2^2} e^{2\tanh \eta\, \left( u_1u_2 - v_1 v_2\right)}. \label{husimi}
\end{align}
Again, in the last step we have expressed it explicitly in terms of the real and imaginary parts of $z_i$. This Husimi function may be factorized as the product of a function of the $(u_1,v_1)$ variables and another of the $(u_2,v_2)$ variables provided that $\tanh \eta = 0$, corresponding to the non-coupled case, i.e without entanglement. This is the case of the composition of multiple Husimi functions in the ground state \eqref{husimiGS}. However, any finite coupling will introduce a certain degree of entanglement.

This state satisfies the normalization condition, in which $\text{tr}(\rho)=1$,
\begin{equation}
    \text{tr}(\rho) = \int F^H_{\Omega_w} \mathrm{d} z_1 \mathrm{d}z_2 = 1.
    \label{traceFullSys}
\end{equation}
We can calculate the trace for the squared density as,
\begin{equation}
    \text{tr}(\rho^2) = (2\pi)^2 \int \left(F^H_{\Omega_w}\right)^2 \mathrm{d} z_1 \mathrm{d}z_2 = \text{sech}^2 \eta,
\end{equation}
where the $(2\pi)^d$ factor (with $d=2$) emerges due to the normalization structure we are considering \cite{NozCoupled}, as we are integrating over the $d=2$ dimensional complex space. We see how, for the decoupled case $\eta=0$, we have a pure state, having a mixed full state otherwise.
\\

It is possible to define a marginal distribution of one of the subsystems in an analogous way to that of the density matrix formalism in the configuration space by integrating the Husimi function with respect to the variables of the other subsystem:
\begin{equation}
	F^{H}_1 (z_1) \equiv \int F^{H}(z_1,z_2)\, \mathrm{d}u_2 \mathrm{d}v_2.
\end{equation}

In this case, we have chosen to integrate in the $(u_2,v_2)$ variables, which yields a distribution for the $(u_1,v_1)$ subsystem. This integral is straightforward in the case of the ground state since it is just a Gaussian-type distribution. The Husimi function of the ground state for the subsystem 1 is:
\begin{align}
	F^{H}_1 (z_1) &= \int F^H_{\Omega_w}(z_1,z_2)\,\mathrm{d}u_2 \mathrm{d}v_2 \nonumber \\
	&= \frac{\text{sech}^2\eta}{\pi}\, e^{-\text{sech}^2\eta\; \left(u_1^2+v_1^2\right)}. \label{husimi1}
\end{align}

If we perform the integral in the variables of subsystem 1, we obtain one, since this is equivalent to calculating the trace of the density operator\eqref{traceFullSys}:
\begin{equation}
	\text{tr} ( \rho ) \equiv \int F^{H}_1 (z_1)\, 
	\mathrm{d}u_1 \mathrm{d}v_1 = 1.
\end{equation}

The trace of the square of the density matrix gives us information about whether the state is a pure or mixed state. In our case for the first subsytem,
\begin{equation}
	\text{tr}(\rho^2) \equiv 2\pi \int  (F^{H}_1 (z_1))^2\,  \mathrm{d}u_1 \mathrm{d}v_1   = \text{sech}^2 \eta\,.
\end{equation}
By definition, a pure state satisfies $\text{tr}(\rho^2)=1$, whereas $\text{tr}(\rho^2)< 1$ holds for mixed states. In the case of our ground state, the subsystem is pure if $\eta =0$ which means that we would just have two decoupled harmonic oscillators, such that measures in one of them does not affect the other one. Then the Husimi function of such state is pure. Whenever there is a coupling within the subsystems $\eta\neq 0$, the state becomes mixed with both oscillators being one indivisible system. \\


As it is well-known, the original ground state is that the expected value of $N_{z_1}$ or $N_{z_2}$ vanishes for $\Omega_z$ (in $\mathcal{Z}$), but not for $\Omega_w$ (in $\mathcal{W}$), and vice-versa for the transformed ground state and the expected value of$N_{w_1}$ or $N_{w_2}$. We can see it by applying the number operator for one of the subsystems to the ground state $\Omega_w$:
\begin{equation}
    N_{z_1} \Omega_w \equiv z_1 \bar{z}_1\, \Omega_w = z_1 \partial_{z_1} \Omega_w =  \tanh \eta\; z_1 z_2\, \Omega_w. \label{op_numero}
\end{equation}

Since the ground state is invariant under the transformation $z_1 \leftrightarrow z_2$, the occupation number for both subsystems is the same. The expected value of the number operator $N_{z_1}$ in the ground state is then:
\begin{align}
    \langle N_{z_1}\rangle_{\Omega_w} &= \frac{\text{sech}^2\eta}{\pi^2}\int\mathrm{d}z_1 \mathrm{d} z_2 e^{ -\abs{z_1}^2-\abs{z_2}^2} \nonumber \\
    & \quad \cdot (\tanh \eta\; z_1 z_2)  \, e^{\tanh \eta\; ( z_1 z_2 + z^*_1 z^*_2)} \,  \nonumber\\
    &= \sinh^2  \eta.
    \label{expvalueN}
\end{align}
This result depends on the coupling parameter $\eta$. For mixed and entangled states ($\eta\neq 0$) we get a non-zero occupation number, however for pure and uncoupled states ($\eta=0$) the expected value is zero. 

\subsubsection{Heisenberg's uncertainty principle}

Heisenberg's uncertainty principle is one of the main features of quantum mechanics. In this section we shall compute the uncertainty associated to the ground state of our system and, in particular, we shall show that the transformed ground state satisfies the uncertainty principle. We will focus on subsystem 1, but this holds for both subsystems. We shall start by calculating the expected value of the position $x_1$.
The Segal-Bargmann position operator acting on the ground state reads
\begin{equation}
	x_1 \Omega_w = \frac{1}{\sqrt{2}} (z_1 + \partial_{z_1})\, \Omega_w =  \frac{1}{\sqrt{2}}(z_1 + \tanh \eta\; z_2 )\, \Omega_w.
\end{equation}
However, the expected value of $z_i$ in the ground state is zero:
\begin{align}
	 \langle z_i \rangle_{\Omega_w} = \frac{\text{sech}^2\eta}{\pi^2}\int\mathrm{d}z_1 \mathrm{d} z_2  e^{-\abs{z_1}^2-\abs{z_2}^2}& \nonumber \\
	 \cdot e^{ \tanh \eta\; ( z_1 z_2 + z^*_1 z^*_2)} z_i \, 
	 = 0,&
\end{align}
so $\langle x_1 \rangle_{\Omega_w}=0$.

In a similar way, the expected value of the momentum is provided by the integral of the following function:
\begin{equation}
	p_1 \Omega_w = \frac{i}{\sqrt{2}} (z_1 - \partial_{z_1})\, \Omega_w = \frac{i}{\sqrt{2}} (z_1 - \tanh \eta\; z_2 )\, \Omega_w,
\end{equation}
which again translates into a zero expected value: $\langle p_1 \rangle_{\Omega_w}  = 0$.

On the other hand, the quadratic term for the position takes the form:
\begin{align}
	x_1^2 \Omega_w &= \frac{1}{2} \left( z_1^2 + \partial_{z_1}^2 + 2 N_{z_1} + 1 \right) \Omega_w \nonumber \\
	&=\frac{1}{2} (z_1^2 + \tanh \eta^2 z_2^2 +2 N_{z_1} + 1) \Omega_w.
\end{align}
In this case, we have a combination of various expected values. The expected value of the number operator is given by \eqref{expvalueN} and that of the identity operator is trivially $1$ \eqref{normalización}. On the contrary, the expected value of $z_i^2$ is zero for both $i=1$ and $i=2$:
\begin{align}
    \langle z_i^2 \rangle_{\Omega_w} =\frac{\text{sech}^2 \eta }{\pi^2}\int \mathrm{d}u_1\mathrm{d}u_2\mathrm{d}v_1\mathrm{d}v_2 e^{ - u_1^2 - v_1^2 } e^{-u_2^2 - v_2^2}& \nonumber \\
    \cdot e^{2\tanh \eta\;\left( u_1u_2 - v_1 v_2\right)} (u_i^2 - v_i^2 + 2i u_i v_i)  = 0.& \label{integra_cuadr}
\end{align}
The $u_1^2$ y $v_1^2$ terms cancel each other and the $u_1v_1$ term is identically zero. Hence, we obtain:
\begin{equation}
	\langle x_1^2 \rangle_{\Omega_w} = \langle N_{z_1} \rangle_{\Omega_w} + \frac{1}{2} = \sinh^2 \eta + 1/2.
\end{equation}

A similar calculation for the momentum results in:
\begin{align}
	\langle p_1^2 \rangle_{\Omega_w} &= \frac{1}{2} \left(- \langle z_1^2\rangle - \tanh \eta^2 \langle z_2^2\rangle +2 \langle N_{z_1}\rangle + 1 \right) \nonumber \\
	& = \sinh^2 \eta + 1/2.
\end{align}

Finally, the uncertainty relation for the position and momentum of the subsystem 1 reads:
\begin{align}
	\Delta x_1 \Delta p_1 &
	= 1/2 + \sinh ^2 \eta \geq 1/2. \label{ppio_indet}
\end{align}

The minimum uncertainty allowed is $1/2$, which is satisfied in the case $\eta=0$, i.e uncoupled and non-entangled oscillators. The uncertainty begins to increase as the oscillators become more strongly coupled and entangled. The bigger the entanglement (bigger $\eta$) the bigger the loss of information in position and momentum (simultaneosly), resulting in a less localized system in phase space. 

The calculation for subsystem 2 is completely analogous and satisfies the same relation as subsystem 1, given by Equation \eqref{ppio_indet}. It means that as $\eta$ increases, both subsystems lose information individually, and this is transferred to the total system overall.

\subsubsection{Correlations}
As we showed, both modes $z_1$ and $z_2$ are entangled over the vacuum state $\Omega_w$. Thus, correlations must exist on measures done among the oscillators. Measuring correlations between two measures is carried by the correlation function,
\begin{equation}
	\mathcal{C}(A,B) \equiv \langle A B\rangle - \langle A\rangle \langle B \rangle.
\end{equation}

If the correlation function $\mathcal{C}(A,B)$ is zero, then both operators $A$ and $B$ are completely independent. On the contrary, if this function is not equal to zero, both measures are correlated. Knowing if these correlations are quantum or classical in nature is not possible from this function. Later, we will see a way to assure that these correlations are purely quantum and, thus, define an entangled system.

We can compute the correlation between two modes in the vacuum state as,

\begin{equation}
	\mathcal{C}(z_1,z_2) = \langle z_1 z_2 \rangle - \langle z_1 \rangle \langle z_2 \rangle = \langle z_1 z_2 \rangle,
\end{equation}

where we have used $\langle z_i\rangle_{\Omega_w}=0$ for all the oscillators. Therefore, the correlation is
\begin{align}
	C(z_1,z_2) =& \frac{\text{sech}^2 \eta }{\pi^2 } \int  \mathrm{d}u_1\mathrm{d}u_2\mathrm{d}v_1\mathrm{d}v_2 e^{ - u_1^2 - v_1^2 -u_2^2 - v_2^2} \nonumber \\
	&\cdot \left[ u_1 u_2 - v_1 v_2 + i ( u_1 v_2 + u_2 v_1) \right]  \nonumber \\
	& \cdot e^{\tanh \eta\; \left( u_1u_2 - v_1 v_2\right)} \nonumber \\
	=& \frac{1}{2} \sinh (2\eta).& \label{val_esp_modos}
\end{align}
We can see how both coupled oscillators are correlated when coupled ($\eta\neq 0$). To know how physical observables, like position and momentum, are correlated, we calculate their correlation function. For the position,
\begin{equation}
	\mathcal{C}(x_1,x_2) = \langle x_1 x_2 \rangle ,
\end{equation}
since $\langle x_i\rangle =0$. The expected value for the positions product $x_1 x_2$ is obtained by applying their corresponding operator on the vacuum state,
\begin{align}
	 x_1 x_2 \Omega_w &= \frac{1}{2}(z_1 + \partial_{z_1} )(z_2 + \partial_{z_2}) \Omega_w \nonumber\\
	 &=  \frac{1}{2}\bigl( z_1 z_2 ( 1+ \tanh^2  \eta)  \nonumber \\
	 &\,  + \tanh \eta ( z_1^2 + z_2^2) + \tanh \eta \bigr) \Omega_w,
\end{align}
and integrating over the measure,
\begin{align}
	\langle x_1 x_2 \rangle  &= \frac{1}{2}\bigl( \langle z_1 z_2 \rangle( 1+ \tanh^2  \eta) \nonumber\\
	&\, + \tanh \eta \langle( z_1^2 + z_2^2)\rangle + \tanh \eta \bigr).
\end{align}
The expected values for the squared terms $z_i^2$ are zero \eqref{integra_cuadr}, whereas the expected value for the  $z_1 z_2$ product was already computed in \eqref{val_esp_modos}. The computation for the momenta correlation is analogous. In particular,
\begin{align}
	p_1 p_2 \Omega_w &= -\frac{1}{2}\bigl( z_1 z_2 ( 1+ \tanh^2  \eta) \nonumber \\
	& \, - \tanh \eta ( z_1^2 + z_2^2)+\tanh \eta\bigr) \Omega_w.
\end{align}
Therefore, the expected value for the correlation between momenta has the same absolute value that the one between the positions but with opposite sign. We can write these correlations as
\begin{equation}
	\mathcal{C}(x_1,x_2) = -\mathcal{C}(p_1,p_2) =   \sinh \eta\; \cosh \eta.
\end{equation}
These correlations show that the value of the position and momentum of one of the subsystems is not independent (if $\eta \neq 0$) of the value of the position and momentum of the other subsystem.

\subsection{Husimi function of coupled harmonic oscillators}

The complete Husimi function from our coupled system is represented in a 5-dimensional space. In order to show different properties of the Husimi quasiprobability distribution, we can choose to fix different variables of its 4-dimensional domain. We are interested in visualizing subsystem 1, so we can fix the $(u_2,v_2)$ variables to eliminate these two dimensions and represent the remaining $(u_1,v_1)$ variables in a 3D plot.

\begin{figure*}[ht]
    \centering
    \includegraphics[trim = {2cm, 2cm, 2cm, 2cm},  width=0.9\textwidth]{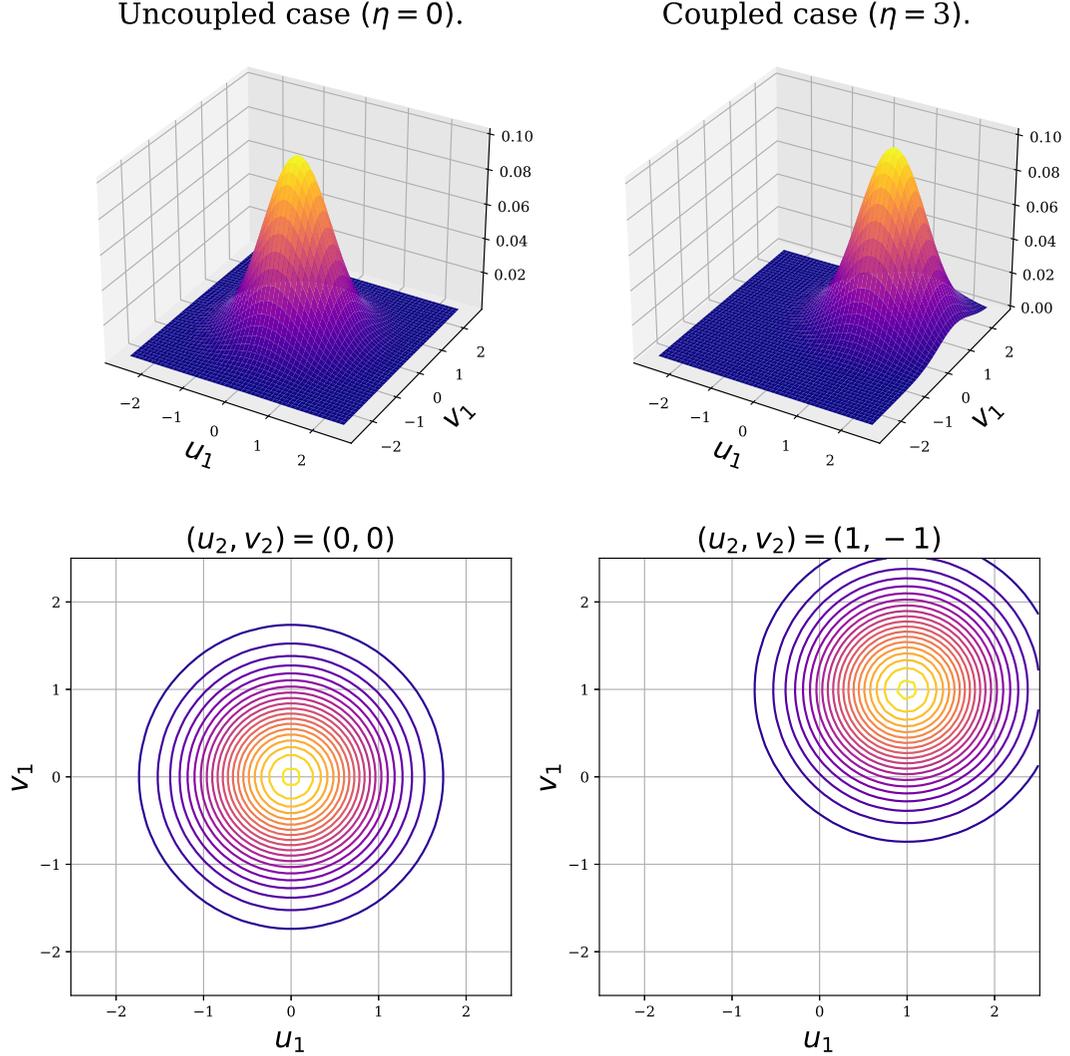}
    \caption{Total Husimi function depicted in real and imaginary parts of subsystem 1. On the left panel, the Husimi function for uncoupled systems $\eta=0$. We represent the case for $(u_2, v_2) = (0,0)$, however the overall shape of the function is always the same (but the scale, for normalization conditions). We see how we find the Husimi function for the ground state of one harmonic oscillator without being affected by the second subsystem. On the right panel, we plot the coupled case with $\eta = 3$. In this case, we fix $(u_2, v_2) = (1,-1)$. One can see that there are obvious correlations between positions and momenta, so the oscillators are not independent.}
    \label{fig:HusimiOA}
\end{figure*}

When both subsystems are uncoupled $(\eta=0)$, both oscillators are independent, thus, for every fixed pair $(u_2,v_2)$, the total Husimi function has the same shape as that of an independent harmonic oscillator but scaled by normalization. When we induce coupling by setting $\eta \neq 0$, the Husimi function moves over the phase space origin of subsystem 1, showing a correlation among subsystems. For instance, for the case of the ground state, for $\eta = 3$, $\mathcal{C}(x_1,x_2) = -\mathcal{C}(p_1,p_2) =\sinh{3}\cosh{3}\simeq 101$. In \autoref{fig:HusimiOA}, we can see that for the  $(u_2,v_2)= (1,-1)$ section, the function maximum changes, which can be thought of as an indicator of a larger probability of finding subsystem 1 closer to the  $(u_1,v_1) \simeq (\tanh{3},\tanh{3})\simeq (1,1)$ position in phase space, given the quasiprobability interpretation of the Husimi function.

The act of fixing points on subsystem 2 does not imply measuring, it is just a way of choosing a subspace of our system, which is correlated. Therefore, entanglement is not a consequence of measurement, but rather a consequence of the probabilistic nature of quantum mechanics. Then, the entanglement properties are within the system independently of measuring it. This gives a satisfactory explanation to the causal entanglement problem, as a measurement is not affecting a subsystem casually disconnected from another.

\subsubsection{Husimi function of the general excited state}

The most general form of an excited state was shown in equation \eqref{excited}. We will call such state $f_{(n_1,n_2)}$, meaning we have subsystem 1 in the $n_1$ excited state and subsystem 2 in the $n_2$ excited state. Taking into account \eqref{op_w} and \eqref{vacio_w} we can first calculate $f_{(n_1,0)}$ in terms of the $z_1$ and $z_2$ variables just by repeated action of the operator $w_1=\cosh{\eta}\,z_1-\sinh{\eta}\, \partial_2$ on $\Omega_w$. Something important to notice is that, once again, the symmetry under the change $z_1 \leftrightarrow z_2$ ultimately means that we can either calculate $f_{(n_1,0)}$ or $f_{(0,n_2)}$ and the result will be the same under this transformation. 

Choosing $n_2=0$ in \eqref{excited} we get:
\begin{eqnarray}
   f_{(n_1,0)}&=&\frac{w_1^{n_1}}{\sqrt{n_1!}}\Omega_w
    \;\;\;\;\;\;\;\;\;\;\;\; \;\;\;\;\;\;\;\;\;\;\;\; \;\;\;\;\;\;\;\;\;\;\;\;
    \;\;\;\;\;\;\;
    \nonumber \\
    &=&\frac{1}{\sqrt{n_1!}}(\cosh{\eta}\,z_1-\sinh{\eta}\,\partial_2)^{n_1}\Omega_w.
\end{eqnarray}

To calculate this, we expand $w_1^{n_1}$ using the binomial theorem, and after some manipulation we obtain:
\begin{eqnarray}
    f_{(n_1,0)}&=&\frac{1}{\sqrt{n_1!}}\sum_{k=0}^{n_1}(-1)^k\binom{n_1}{k}
     \nonumber \\
    & &\tanh^k{\eta}\,\cosh^{n_1}{\eta}\;
     z_1^{n_1-k}\,\partial_2^k\Omega_w\,.
\end{eqnarray}

The k-th derivative of $\Omega_w$ with respect to $z_2$ is easily found to be simply $(\tanh{\eta}\,z_1)^k\,\Omega_w$, so we can finally compute our desired function:
\begin{align}
    f_{(n_1,0)}&=\frac{(\cosh{\eta}\,z_1)^{n_1}}{\sqrt{n_1!}}\Omega_w\sum_{k=0}^{n_1}(-1)^k\binom{n_1}{k}\tanh^{2k}{\eta} \nonumber \\
    &=\frac{\text{sech}^{n_1}\eta}{\sqrt{n_1!}}z_1^{n_1}\Omega_w\,.
    \label{n1,0}
\end{align}

Now, all that is left for us to calculate $f_{(n_1,n_2)}$ is to see what is the action of the operator $w_2=\cosh{\eta}\,z_2-\sinh{\eta}\, \partial_1$ on our previous result \eqref{n1,0}. The strategy is again to expand $w_2^{n_2}$ by means of the binomial theorem. In a similar way as before, apart from the summation and some hyperbolic factors, it will appear the k-th derivative of $z_1^{n_1}\Omega_w$ with respect to $z_1$ this time. This requires us to use the generalized Leibniz rule:
\begin{equation}
    \partial_1^{k}(z_1^{n_1}\Omega_w)=\sum_{l=0}^{k}\binom{k}{l}\partial_1^{k-l}(z_1^{n_1})\partial_1^{l}(\Omega_w).
\end{equation}

We notice that $\partial_1^{l}(\Omega_w)$ is already known to be $(\tanh{\eta}z_2)^{l}\Omega_w$. The other derivative is simply the derivative of a power of $z_1$ with respect to $z_1$ itself, so we can write this as:
\begin{align}
    \partial_1^{k-l}z_1^{n_1}&=n_1(n_1-1)...(n_1-k+l+1)z_1^{n_1-k+l} \nonumber \\
    &=(n_1)_{k-l}\,\,z_1^{n_1-k+l},
\end{align}
where $(n_1)_{k-l}$ is the falling factorial. After some manipulation we can write:
\begin{align}
    f_{(n_1,n_2)}=
    &\frac{(\cosh{\eta}\,z_2)^{n_2}}{\sqrt{n_2!}}f_{(n_1,0)} \notag \\
    & \cdot \sum_{k=0}^{n_2}(-1)^k\binom{n_2}{k}\left(\frac{\tanh{\eta}}{z_2z_1}\right)^{k} \nonumber \\
    &\cdot \sum_{l=0}^{k}\binom{k}{l}(n_1)_{k-l}\,(\tanh{\eta}\,z_2z_1)^{l}\,,
\end{align}
and once the summations are performed:
\begin{eqnarray}
    & &f_{(n_1,n_2)} = f_{(n_1,0)} \frac{z_2^{n_2} }{\sqrt{n_2!}} (\cosh{\eta})^{n_2}  h(n_2+1) 
    \;\;\;\;\;\;\;\;\;\;
    \nonumber\\
    & & \,
    =\frac{z_1^{n_1}}{\sqrt{n_1!}}\frac{z_2^{n_2}}{\sqrt{n_2!}} \left(\cosh \eta \right)^{(n_2-n_1)} h(n_2+1)\,\Omega_w, \label{generalState}
\end{eqnarray}
with $h(n)$ the recursive relation (holonomic sequence):
\begin{align}
   &h(n+2)  \bigg[\tanh \eta\;  (n-n_1+1) (n-n_2+1)  \nonumber  \\
    &+ z_1 z_2 \tanh^2\eta\;(n_2-n - 1) - (n+2) z_1 z_2 \bigg]  \nonumber \\
    &-h(n+1) \tanh\eta\;(-n+n_2-1)  \notag \\
    &\left[\tanh^2\eta\;(n_2-n)-n +n_1+z_1 z_2 \tanh\eta\;-1\right] \nonumber \\
    &+ h(n) \tanh^3\eta\;(n_2-n-1) (n_2-n) \nonumber \\
    &+ h(n+3)(n+2) z_1 z_2  = 0\, ,
\end{align}
with the initial values,
\begin{align}
    h(0) &= 0, \quad h(1) = 1, \\
    h(2) &= 1 - n_2 \left[\tanh^2\eta\; + \frac{n_1 \tanh\eta}{z_1 z_2}\right].
\end{align}

The Husimi function cannot be analytically calculated unless $n_1$ and $n_2$ are specified due to the $h(n_2+1)$ recursive relation. However, in a symbolic way, we can express it as:
\begin{align}
    F^H_{(n_1,n_2)}&=\abs{f_{(n_1,n_2)}}^2\mu(z_1,z_2) \nonumber \\
    &=\frac{(\cosh{\eta})^{2(n_2-n_1)}}{\pi^2\,n_1!n_2!}h^2(n_2+1)\,\abs{z_1^{n_1}}^2\abs{z_2^{n_2}}^2\, \nonumber\\
    & \, \abs{\Omega_w}^2e^{-\abs{z_1}^2-\abs{z_2}^2},
\end{align}
 which, in terms of its real and imaginary parts reads

\begin{align}
    F^H_{(n_1,n_2)}=&\frac{(\cosh{\eta})^{2(n_2-n_1-1)}}{\pi^2\,n_1!n_2!}\, (u_1^2+v_1^2)^{n_1}(u_2^2+v_2^2)^{n_2} \notag \\
    & e^{-(u_1^2+u_2^2+v_1^2+v_2^2-2\tanh{\eta}\,(u_1u_2-v_1v_2))} \nonumber \\
    & h^2(n_2+1) \,.
\end{align}

\section{Wehrl entropy}\label{sec:entropy}

The Husimi function can be considered as the Segal-Bargmann operators' representation of the density matrix. Then, this function is a representation of the state in a space equivalent to phase space. Thus, an entropy can be defined for this density, the Wehrl entropy \cite{wehrl}:
\begin{equation}
\label{Wehrl}
	S^W = - \int F^H(z) \ln F^H(z) \mathrm{d}z,
\end{equation}
where $z$ generally accounts for the total number of coordinates $z_1,z_2,...$ associated with the total SBS. Given that we use natural units setting $k_B=1$ and $\hbar =1$, the integral in the entropy is not dimensionless, as it should be. To recover physical units, the $1/\hbar^d$ factors must be included in the Husimi function. Therefore, the physical entropy is  $ k_B \ln(\hbar^d)\hbar^d $ times the dimensionless entropy, with $d$ the dimension of the system. 
\\

Wehrl entropy is particularly well suited for our calculations in Segal-Bargmann space, since it is defined in terms of the Husimi function. It is a quasi-classical entropy that measures the loss of information in phase space. It is not fully classical since it cannot be negative, for it must satisty the relation: $S^W\geq 1$. Another important property is that, if $S$ is the standard quantum entropy $S = - \mathrm{tr} \rho \ln \rho$, then this entropy gives a lower bound to Wehrl entropy as $S \geq S^W$. As Wehrl showed \cite{wehrl}, the quasi-classical entropy is a good approximation as long as the Husimi function is a smooth function spread over a phase space volume bigger than $h$. Thus, the Wehrl entropy is a good approximation for our harmonic oscillators. Otherwise, for exact location in phase space systems, i.e. a Dirac Delta in position like the EPR thought experiment, the quasi-classical approximation has problems.

As we are working with a system in a Hilbert space $\mathcal{H}$ that we split into two subspaces $\mathcal{H} \equiv \mathcal{H}_1 \otimes \mathcal{H}_2$ corresponding to each subsystem, we can calculate the partial Husimi function for one subsystem by integrating the other subsystem variables $F^H_1 = \mathrm{tr}_2 F^H$. This partial density has an entropy associated, that we will name as partial entropy $S^W_1$.
For a bipartite system, this entropy of one of the subsystem is called the entropy of entanglement. For a pure total state, the entropy of the subsystems can quantify the entanglement of the bipartite states. This entropy can be interpreted as the maximal amount of classical information that measuring one system can provide about the results on measurements performed on the second one \cite{janzing}. For the Husimi distribution, the Wehrl entropy of entanglement is monotonous under partial trace. Also, the total entropy gives an upper-bound to the entropy of entanglement as $S^W_1 \leq S^W$.\\

The partial entropies of both subsystems and the associated to the total system allow us to calculate the correlation between the two subsystems. From Wehrl entropy, we calculate Wehrl mutual information \footnote{This expression holds as long as each term is finite. Clearly, this is our case for finite coupling $\eta$.},
\begin{equation}
	I^W(\rho_1:\rho_2) \equiv S^W_1 +S^W_2 - S^W, \label{mutual}
\end{equation}
where $\rho_1$ and $\rho_2$ correspond to the partial density operators of subsystems $1$ and $2$ respectively, obtained from the total $\rho$ of the whole system \cite{Floer}. It is always positive, except when the state $\rho$ is a product state, which is 0. Mutual information can not distinguish between classical and quantum correlation. However, if $\rho$ is pure, then the correlations showed in the mutual information are of quantum nature. Then, for pure states, the mutual information is a quantum entanglement witness.

In the next part, we compute the Wehrl entropy for our system, calculating the entropy for the ground state. Afterwards, we show a way to generalize this computation for any excited state and as, an example, we apply it to different excited state.

\subsection{Wehrl entropy for the ground state}
The Husimi function of the ground state \eqref{husimi}, which we will refer to as $F_{(0,0)}^H$, can be written as $\mathcal{N}\exp [\text{Arg}(z)]$.  Thus, the Wehrl entropy can be obtained efficiently as,
\begin{align}
    S^W &= - \int F^H(z) \ln \left( \mathcal{N} e^{\text{Arg}}\right) \mathrm{d}z \nonumber \\
	&= \ln \mathcal{N}^{-1} - \mathcal{N} \int e^{\text{Arg}} \text{Arg} \, \mathrm{d}z, \label{Wehrl2}
\end{align}
where, in the first term, we used the fact that the Husimi function is properly normalized. This expression allows us to obtain the Wehrl entropy just from an integral. This kind of Gaussian integral is done for the rest of the calculations, so from now on, they will be skipped.

By taking the Husimi function of the ground state system,  we can compute the total entropy by using the same procedure with the total Husimi function. Again, calculation is straightforward using Gaussian integrals, with total entropy,
\begin{equation}
	S^W_{(0,0)} = 2 + 2\ln (\pi) +2\ln{(\cosh \eta)} \,.
\end{equation}
As we see, the Wehrl entropy is always positive and greater than a minimum value $S^W_{(0,0)} \geq 2 \left( 1 + \ln{( \pi)}\right)$. The entropy increases as the coupling parameters does. So, when there is not coupling ($\eta = 0$), the entropy corresponds to that of a set of 2 harmonic oscillators in the ground state. The factor $2$ in the minimum corresponds to the number of oscillators, as the $\ln \pi$ term is associated with normalization for a bipartite system and the $2$ term is the number of subsystems\cite{Floer}. Thus, we can generalize the result for $N$ uncoupled oscillators in the ground state being $S^W_{N(0)} = N \left( 1 + \ln{(\pi)}\right)$.
\\

We can also calculate it by considering the marginal distributions for each subsystem individually. For subsystem 1, we use Equation \eqref{husimi1} to get,
\begin{align}
	S^W_{1 (0,0)} &=   \frac{\text{sech}^2\eta}{\pi} \int e^{-\text{sech}^2\eta\left(u_1^2+v_1^2\right)}  \nonumber \\
	& \left[ - \text{sech}^2 \eta\;(u_1^2 + v_1^2)\right] \mathrm{d}u_1 \mathrm{d}v_1 + \ln ( \pi \cosh^2 \eta ) \nonumber \\ 
	&= \ln (\pi) +  2\ln (\cosh\eta) + 1.
\end{align}

Because of the symmetry of the vacuum state under the change $z_1 \leftrightarrow z_2$, the partial Husimi function from subsystem 2 coincides with the one from subsystem 1. Thus, the entropy of each subsystem is the same
\begin{equation}
	S^W_2 = S^W_1 = 1+ \ln (\pi) +  2\ln (\cosh\eta).
\end{equation}
As discussed before, when there is not coupling, the entropy is $S^W_1 ( \eta = 0) = 1 + \ln {(\pi})$, corresponding to one harmonic oscillator. As the coupling increases, the entropy does too. It behaves the same way as the total entropy, growing with the monotonous function $\ln{(\cosh \eta)}  \geq 0$ that secures the positivity of the Wehrl entropy.
\\

\begin{figure*}[ht]
    \centering
    \includegraphics[width=0.9\textwidth]{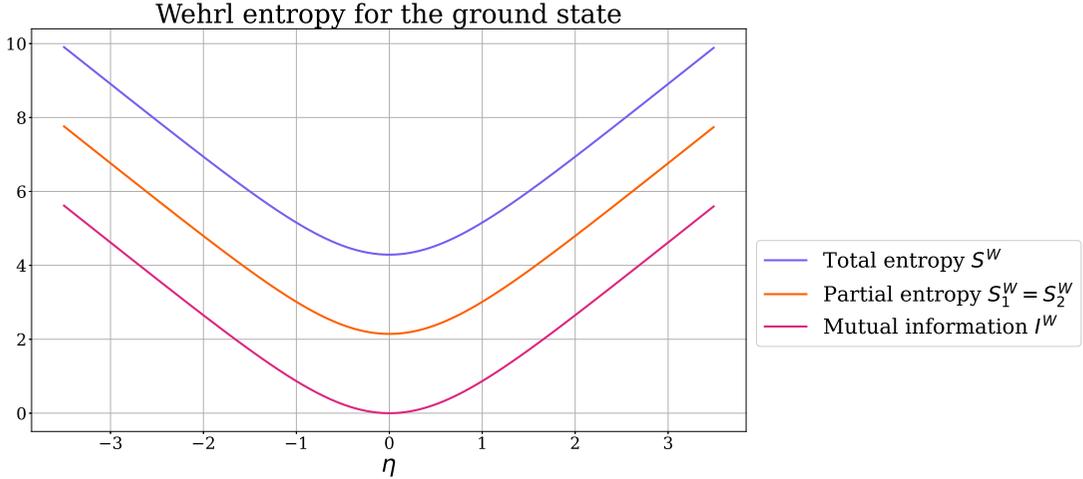}
    \caption{Wehrl entropy for the ground state $(0,0)$ of the system formed by two harmonic oscillators in terms of $\eta$ (a parameter directly related with the coupling between the oscillators). In particular, we plot the total entropy, partial entropy and the mutual information between the oscillators.}
    \label{fig:ground}
\end{figure*}

Now, we calculate the mutual information \eqref{mutual} for the ground state,
\begin{equation}
	I^W_{(0,0)} = 2\ln (\cosh\eta),
\end{equation}
which shows that correlations between both subsystems are purely quantum if $\eta \neq 0$. These correlations do not have a classical equivalent, and can only be a consequence of the entanglement of the vacuum state. We represent this results for the ground state in Figure \ref{fig:ground}.

\subsection{Entropy of the excited states of one of the subsystems}

We now calculate the entropy of the states generated by exciting the subsystem $w_1$ while leaving the subsystem $w_2$ at ground level.  
These states have the general form:
\begin{equation}
    f_{(n,0)} = \frac{w_1^n}{\sqrt{n!}} \Omega_w.
\end{equation}
Note, however, that the symmetry under the change $z_1 \leftrightarrow z_2$ in \eqref{op_w} means that the state ($n,0$) is no different from the state ($0,n$). We will work with the ($n,0$) state, but keep in mind that the results are the same that in the ($0,n$) case. \\

The Husimi function for the ($n,0$) excited state can be written as
\begin{align}
    F^H_{(n,0)}&=\frac{\text{sech}^{2(n+1)} \eta }{\pi^2 n!}(u_1^2 + v_1^2)^{n}  \nonumber \\
    & \,e^{  - u_1^2 - u_2^2 - v_1^2 - v_2^2 + 2 \tanh \eta\; (u_1 u_2 - v_1 v_2)}.
\end{align}

We can use Equation \eqref{Wehrl2}  to compute the Wehrl entropy associated to this Husimi function,
\begin{align}
    S^W_{(n,0)} = & 2 \left[ 1 + \ln{( \pi)} + \ln{( \cosh \eta)} \right] \nonumber \\
    & +n \left( 1 + \gamma - H_n\right) + \ln{( n!)},
\end{align}
where $H_n$ is the $n_{th}$-harmonic number and $\gamma$ is the Euler-Mascheroni constant.

From this result, we can obtain the entropy for the ground state ($n=0$) calculated separately in the previous subsection. The case $n=1$ yields the entropy for the first excited state ($1,0$),
\begin{equation}
	S^W_{(1,0)} = 2 \left[1 + \ln{( \pi)} + \ln{( \cosh \eta)} \right] + \gamma.
\end{equation}

The relative entropy between the $(n+1,0)$ and $(n,0)$ state is,
\begin{equation}
    S^W_{(n+1,0)} - S^W_{(n,0)} =\gamma -H_n + \ln{ (n+1)}.
\end{equation}
As we can see, in the limit $n \to \infty$, the relative difference of entropies tends to zero, since $\text{lim}_{n\to \infty} \left(\ln{( n)} - H_{n-1}\right) = -\gamma $. However, the entropy of that state tends to infinite, as expected. \\
\\

We follow up by calculating the entropy associated with the subsystems using the Wehrl entropy over the partial Husimi functions. The calculations are similar to the full entropy case. The Husimi function for subystem 1 reads
\begin{align}
    F_{1(n,0)}&= \int F_{(n,0)} \mathrm{d}z_2  \nonumber \\
    &= \frac{\text{sech}^{2(n+1)} \eta }{\pi n!} e^{-(u_1^2 + v_1^2)\text{sech}^2 \eta } (u_1^2 + v_1^2)^n. 
\end{align}

It means that its corresponding entropy is
\begin{align}
    S^W _{1(n,0)} &= 1 + \ln{( \pi)} +2 \ln{( \cosh \eta )} \nonumber \\ 
    &\, + n\left(1+ \gamma -H_n\right) + \ln{( n!)}\,.
\end{align}
With this expression, setting $\eta = 0$, we find the general expression for the entropy of one decoupled oscillator at level $n$,
\begin{equation}
    S^{W,dec} _{1(n,0)} = 1 + \ln{( \pi)}  + n\left(1+ \gamma -H_n\right) + \ln{( n!)}. \label{entropyDec}
\end{equation}

For any coupling, the first excited state, $n=1$,
\begin{equation}
    S^W _{1(1,0)} = 1 + \ln{( \pi)} +2 \ln{( \cosh \eta)}  + \gamma.
\end{equation}

Calculating the partial entropy for an arbitrary $n$ is quite challenging, since for $n\geq 2$, the entropy includes a term with a logarithm of a series of terms that depend on all of the variables. In any case, for illustration purposes, we can calculate the entropy for the first excited state ($1,0$). Its partial Husimi function associated with the second subsystem is,
\begin{align}
    F^H_{1(1,0)}(z_2)&=\frac{\text{sech}^{4} \eta }{\pi} e^{-\text{sech}^2 \eta\; (u_2^2 + v_2^2)} \nonumber \\
    & \,\left(\tanh ^2\eta\; \left(u_2^2+v_2^2\right)+1\right),
\end{align}
and its corresponding entropy,
\begin{align}
    S^W_{1(1,0)} &= 1+ \ln{(\pi)} + 4\ln{( \cosh\eta)} \nonumber \\
    & \, - \tanh ^2 \eta \; e^{\text{csch}^2\eta\;} \Gamma \left(0,\text{csch}^2\eta\;\right),
\end{align}
where $\Gamma(s,x)$ stands for the incomplete gamma function. This last expression works well for $\eta \neq 0$. In the case where there is not coupling, the entropy is the same as the ground state for one oscillator \eqref{entropyDec}. Nevertheless, we can obtain the expected result for the entropy if we take the limit $\eta \to 0$. The relative entropy between the first excited state and the ground state is,
\begin{align}
    S^W_{2\,(1,0) } - S^W_{2\,(0,0)} &= 2 \ln{( \cosh \eta)}  \nonumber \\
    & \,- \tanh ^2 \eta \; e^{\text{csch}^2\eta\;} \Gamma \left(0,\text{csch}^2\eta\;\right)\, ,
\end{align}
$\forall \eta \neq 0$ and 
\begin{equation}
    S^W_{2\,(1,0)} - S^W_{2\,(0,0)} = 0 \,,
\end{equation}
for $\eta =0$.
It is interesting to note that this difference is zero when the subsystems are decoupled. In such a case, the excited character of the first oscillator does not affect the entropy of the ground state of the second oscillator. On the contrary, for a finite coupling, the entropy of the second subsystem is increased by the excited character of the first subsystem. Indeed, for very high values of the coupling, in the limits $\eta\to \pm \infty$, the ground state of the second oscillator acquires the same entropy than the one associated with the excited state of the first one,
\begin{equation}
    \lim_{\eta \to \pm \infty}  S^W_{2(1,0) } = 1+ \ln{(\pi)} + 2\ln{( \cosh\eta)} = S^W_{1(1,0)}.
\end{equation}

The mutual information of this state is,
\begin{align}
    I^W(1:2)_{(1,0)} &= 4\ln{( \cosh \eta )} \nonumber \\
    &\, - \tanh ^2 \eta\ e^{\text{csch}^2\eta\;} \Gamma \left(0,\text{csch}^2\eta\;\right),
\end{align}
that can be rewritten as,
\begin{equation}
	I^W_{(1,0)}(\eta)  = I^W_{(0,0)}(\eta) + \mathcal{I} (\eta)_{(1,0)},
\end{equation}
with 
\begin{align}
    \mathcal{I} (\eta)_{(1,0)} & =2\ln{( \cosh \eta )} \nonumber \\
    &\, - \tanh ^2 \eta\ e^{\text{csch}^2\eta\;} \Gamma \left(0,\text{csch}^2\eta\;\right)\, ,
\end{align}
a function that is added because of the excited state. 
Equivalently to the last analysis, it is interesting to study the behaviour of the system for large coupling. In the limit $\eta \to \infty$ this function becomes,
\begin{equation}
    \lim_{\eta \to \pm \infty} \mathcal{I}_{(1,0)} =  \gamma
\end{equation}
so the mutual information for the $(1,0)$ and the ground state, in this limit, differs by the Euler constant $\gamma$. 

\begin{figure*}[ht]
    \centering
    \includegraphics[width=0.9\textwidth]{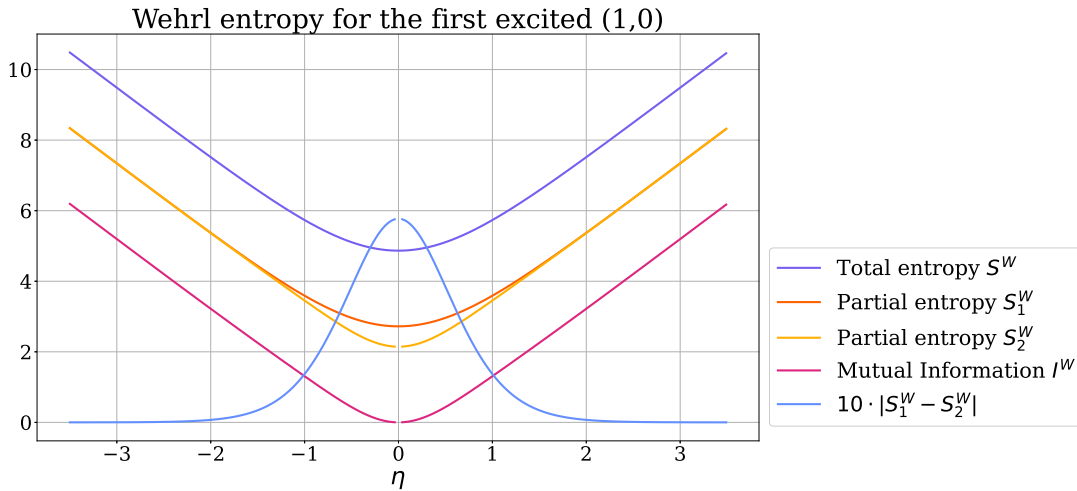}
    \caption{Wehrl entropy for the first excited state ($1,0$) of the system formed by two harmonic oscillators in terms of $\eta$ (a parameter directly related with the coupling between the oscillators). Here, we plot the total entropy, partial entropies and mutual information for this first excited state. In addition, we plot the difference between partial entropies (scaled by a factor 10).}
    \label{fig:entropy10}
\end{figure*}

As we see in Figure \ref{fig:entropy10}, the total entropy has a similar behaviour as the one of the ground state, it is in the partial entropies where we find differences. The partial entropy for the excited first subsystem has a higher value near the decoupled condition than the second subsystem. However, when the coupling starts increasing, both partial entropies get equal. We see this, if we plot the absolute value of the difference between both partial entropies. In the plot (scaled by 10), we see how in the decoupled case, this value is the Euler gamma, and goes to zero in the infinite coupling case.

\subsection{Entropy for excited states of both subsystems.}
We would want to obtain the analytic entropy expression for the general ($n_1,n_2$) state. However, we find some problems to achieve it. First, working with the analytic expression for the Husimi function from the Segal-Bargmann state \eqref{generalState} is impossible, as it has a recursion relation. This first issue can be solved by calculating the Husimi function for given $n_1$ and $n_2$ values, thus obtaining a Husimi function with which we could work. The second problem rises in next step. As we discussed before, obtaining an analytical integration on the entropy for some complex $n$ values is hard, or in some cases, straight up impossible. So, numerical integrations should be carried out if we want entropy for any ($n_1,n_2$) state. However, we stop at the first excited state as we can check the behaviour of the entanglement well enough with this it.

\section{Conclusions}
The objective of this manuscript has been the study of Wehrl entropy in entangled harmonic oscillators. Because entanglement is a purely quantum property, to carry out this study  the first thing we needed was a Hilbert space that would describe our specific problem. Since we were studying a system consistent in harmonic oscillators, it was specially convenient to use the Segal-Bargmann space as our Hilbert space as shown in Section \ref{sec:teo}. 

With the SBS already established, setting up the Hamiltonian of the coupled harmonic oscillator to solve it was straightforward. During the derivation of the eigenfunctions of the problem, the use of the Bogoliubov transformations applied to ladder operators was essential to simplify the calculations since it allowed us to rewrite the Hamiltonian expression with two decoupled terms. Once this expression was reached, the solution of the problem was immediate. In other words, due to how eigenfunctions were constructed, it was only necessary to know the analytical expression of the ground state $\Omega_w$, the remaining eigenfunctions could be obtained using the ladder operators. This last fact allowed us to obtain a general expression for any eigenfunction,
which means that we could calculate the Husimi distribution for any eigenfunction as shown in section \ref{sec:OA}. 

Some remarkable information of the solutions obtained is that they were entangled for the original variables, as expected, and this fact was checked explicitly for the ground state. In addition, it was also verified that the solutions were in accordance with the Heisenberg uncertainty principle. One of the greatest advantages of using the Segal-Bargmann space is that the Husimi quasi-probability distribution emerges naturally. Thanks to the Husimi distribution, physical measurements can be made, however its most important feature for our study, is its direct relation with the Wehrl entropy. Thanks to the Wehrl entropy, it was possible to calculate the physical information and the correlations between variables for the different eigenfunctions of the problem as shown in Section \ref{sec:entropy}. 

On the one hand, we pay special emphasis on the fundamental state and how the entanglement is manifested in it. The ground state has the advantage that all the associated calculations are simple to perform, and from it, the other eigenfunctions arise in a straightforward way by the use of the ladder operators. After a detailed study, we concluded that the ground state had correlations between the two subsystems with no classical analogue, which implies quantum entanglement. On the second hand, we also studied the entropy for the excited states of one of the two subsystems. The results that we obtained were satisfactory, since they had a similar behaviour to those of the ground state.

In summary, we have analyzed the Wehrl entropy of entangled harmonic oscillators. For such a purpose, we have solved the system within the Segal-Bargmann formalism. We have developed this approach successfully for the first time along this work. We plan to extend it for different systems in future analyses, for instance, by studying the introduction of anharmonic effects. 

\bmhead{Acknowledgments}
We are grateful to Ángel Rivas for useful discussions.
This work was partially supported by the MICINN (Spain) project PID2019-107394GB-I00.





\bibliography{ref}


\end{document}